# Power Aware Visual Sensor Network for Wildlife Habitat Monitoring

Mohsen Hooshmand, Shadrokh Samavi, S. M. Reza Soroushmehr
Department of Electrical and Computer Engineering
Isfahan University of Technology, Iran
Isfahan, Iran
Samavi96@cc.iut.ac.ir

*Abstract*— **One of the fundamental issues in wireless sensor network is conserving energy and thus extending the lifetime of the network. In this paper we investigate the coverage problem in camera sensor networks by developing two algorithms which consider network lifetime. Also, it is assumed that camera sensors are distributed randomly over a large area in order to monitor a designated air space for environmental surveillance of the wild life. To increase the lifetime of the network, the density of distributed sensors could be such that a subset of sensors can cover the required air space. As a sensor dies another sensor should be selected to compensate for the dead one and reestablish the initial coverage area. This process should be continued until complete coverage is not achievable by the existing sensors. Thereafter, a graceful degradation of the coverage is desirable. The goal is to elongate the lifetime of the network while maintaining a maximum possible coverage of the designated area. Since the selection of a subset of sensors for complete coverage of the target area is an NP-complete problem we present a class of heuristics for this purpose. This is done by prioritizing the sensors based on their visual and communicative properties.**

*Keywords- visual sensor network; directional sensors; visual coverage; priority based algorithms*

## I. INTRODUCTION

Wireless camera sensor network (WCSN) is constructed by a set of small and low cost sensor nodes which can produce images or videos from the sensing area (Fig. 1). Resource limitations in WCSNs, place numerous constraints on these networks. Limitations on bandwidth, power source, and spatio-temporal sampling capabilities are examples of constraints that need attention in these networks [1]. Generally, sensor nodes operate by batteries and have a limited supply of energy. Therefore, energy consumption has been the focus of different research works [2].

These networks are highly applicable in security surveillance, environment monitoring, industrial process control, object tracking, and person locator services [3], [4], [5]. Another application of such networks is monitoring of wildlife habitat. In these types of applications the monitored area is a vast remote area and large numbers of wireless sensors are randomly deployed by a possible airborne dropdown. Due to their broad applications, these networks have been widely addressed in recent works.

An interesting and versatile application of WCSNs is in security, surveillance and environment monitoring. Therefore, seamless coverage of a vast area is an important issue in these applications. It has been shown that in regular sensor networks, coverage is an NP-complete problem [6]. It is also proved in [7] that the NP-completeness problem exists for the coverage in WCSNs. Hence, heuristics have been proposed for the visual coverage problem for indoor environments such as shopping centers, hospitals, and airports [7], [9], [10], [11]}.

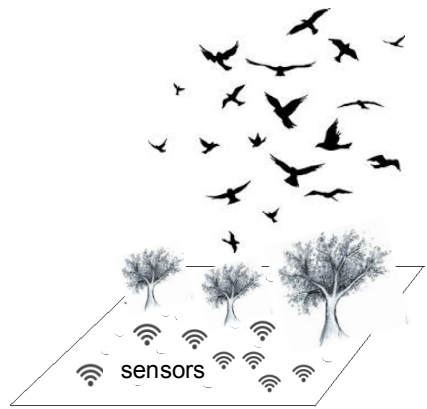

Fig. 1. Environmental monitoring sensor network.

The objective of sensor coverage problems is to conserve energy usage by minimizing the number of active sensors. The selected sensors should form sufficient coverage of the intended region [12]. Various parameters have been considered by different solutions in the literature, including energy efficiency and bandwidth allocation issues. Authors of [13] use distributed look-up tables to rank the cameras according to how accurately they capture the image of a location, and hence chose the best candidates for the image production. Cost metrics can be applied to select a set of cameras that provide a reconstruction of a view from a user-specified view point [11]. Distributed power management of camera nodes based on coordinated node wake-ups is applied by Zamora et al. [14] to lower the energy usage of cameras. The problem of selecting a minimum number of sensors and assigning orientations to the nodes is addressed by Fusco [10]. The orientation assignment is performed such that a given area (or set of target points) is k-covered, so that each point is covered by at least $k$ sensors.

In this paper surveillance of air space through a terrestrial WCSN with randomly distributed sensors is considered. This results in a random location and a random orientation for each sensor node. The density of placement of the sensors is such that only a subset of the sensors would be enough to cover the whole designated air space. Therefore, it would be possible to elongate the lifetime of the surveillance through suitable selection of subsets of cooperating sensors. All of the sensors are initially calibrated. Hence, the location and orientation of each sensor are known to a base station. Each sensor can communicate its visual perceptions to its neighboring nodes. In order to have an optimum coverage of the area with the minimum number of sensor nodes, an appropriate subset of these nodes must be selected at each time, which is an NP-complete task. In this paper, two algorithms are proposed to generate maximum coverage area while the elongation of the lifetime of the network is the goal. Sensors are prioritized for selection process based on a number of characteristics. Parameters such as coverage area, remaining energy in each sensor node, or the overlap of sensors coverage are considered in these algorithms. Through rigorous simulations we evaluated the proposed algorithms. The most efficient algorithm is found through comparison of the obtained results.

The remainder of the paper is organized as follows. In section II the coverage problem in WCSN is formulated and a number of definitions that are used in the rest of the paper are stated. Specific selection algorithms are proposed in section III. Simulation results are provided in section IV. Section V is dedicated to some concluding remarks.

## II. PROBLEM STATEMENT AND CHARACTERISTICS

In this paper, we assume that a WCSN is employed for airspace surveillance. The most important factor which should be considered in the design of this WCSN is the coverage. The network is monitoring a target region, at a certain threshold height, in the air space under surveillance and needs a complete coverage of that region. To do so, the provided images from different cameras can be used to generate a panoramic image in the base station. If the target region is completely covered then space above it is most likely to be covered too. Our main goal is to provide the maximum coverage of the target region with the minimum number of sensor nodes, while maximizing the network lifetime. In this section, we formulated this problem in more details.

Let us assume that sensor nodes are uniformly distributed over the sensing area. Each sensor node has an orientation that can be specified by three different angles with respect to a global coordinate system. Each camera sensor node knows its own coordinates and orientation through an initial calibration phase. The next issue is to determine an optimum coverage of the sensing area. Due to the fact that finding the optimum number of sensors for an appropriate coverage is an NP-Complete problem, sub-optimal techniques must be employed.

Because of the limited source of energy in each sensor node, the number of sensors with adequate energy varies with time. After a sensor operates for a while its battery runs out due to either capturing images or communicating with other sensors. Prior to exhaustion of battery, a sensor is said to be live and after that the sensor becomes dead. Live sensors are employed to monitor an airspace region of area $A$. Each sensor node is at a different coordinate with a different orientation. Hence, each subset of sensors has its own coverage area. The coverage set of a camera sensor is defined as follows:

**Definition (Coverage)** The coverage of a camera sensor is the set of points in the target region that is covered by that sensor. In other words, the intersection of a sensor's field of view (FOV) and the target region is defined as the coverage of that sensor. Let $S$ be an arbitrary set of sensors then we define $C(S)$ to be the coverage set of $S$.

In wireless networks, each mobile station has a transmission range of $r$. Two mobile stations are considered as neighbors if and only if their distance is less than or equal to $r$. Neighbor nodes can directly communicate through a single hop radio channel. When the distance between the source and destination nodes is greater than $r$, multi-hop routes are required. However, in WCSNs, geographically neighbor nodes do not necessarily sense adjacent visual areas. On the other hand, two camera sensors can sense similar parts of the visual environment while their geographical distance could be greater than $r$. Hence, in WCSNs we can consider sensor nodes as neighbors from two different points of views. First, two sensors are visually neighbors if there exists overlap in their FOVs.

Second, two sensors are geographically neighbors if the distance between them is less than their transmission range.

Since there may exist some sensors with overlapped coverage sets, different sensor sets can be selected and hence we should employ a selection algorithm to select a subset of sensors that provides an appropriate total coverage. As was mentioned, sensors are either live or dead. The live ones are either active or inactive. The sensors that are selected, either for visual sensing or routing purposes, are considered as active, while the rest of sensors are either inactive or dead. Let $L_i$ and $S_i$ denote respectively the sets of live and selected sensors at $i^{th}$ time step of the network's operation. Also suppose that each sensor is live at most $k$ time steps. Therefore $i$ varies from 1 to $T = |L_0| * k$ such that $L_0$ is the set of live sensors right at the beginning of network operation and $|L_0|$ represents the cardinal number of $L_0$. Since $S_i$ is a subset of $L_i$ there are $2^{|L_i|}$ choices for constructing $i$ for all $i$ from 1 to T. Our goal is to select a set of sensors at each time step which has some properties defined in the followings.

**1- Definition (IC)** If the coverage of selected sensors from the first step to $a^{th}$ time step is equal to $C(L_0)$ then this sequence of selected sensors (i.e. $S$) has *Initial Coverage stability*, IC, property and $a$ is shown by $a_{IC}(S)$.

It is obvious that after $a_{IC}$ the amount of network coverage decreases. In some applications, the network would be useful until a percentage of initial coverage is achieved. This property is defined in the following.

**2- Definition (FC)** A sequence of selected sensors has the property *feasible coverage*, FC, if there exists $a \leq T$ such that for all $i \leq a, |C(S_i)| \geq \gamma \times |C(L_0)|$. In such a case, parameter $a$ is the maximum time duration of the network usefulness and is represented by $a_{FC}(S)$. Then the achieved coverage is greater than or equal to $\gamma \times |C(L_0)|$ where $\gamma$ is a threshold value between 0 and 1.

Although achieving the complete coverage for the maximum duration is the ultimate goal of the surveillance cases, after a while, with the death of a large number of sensor nodes, this goal will no longer be achievable. Under these conditions, the task is degraded to achieving an acceptable level of coverage, $\delta$. This property is defined in the following.

**3- Definition (AC)** If for a sequence of selected sensors, $S$, there exists $a \geq a_{IC}(S)$ such that for all $i \leq a$ and $i \geq a_{IC(S)}$, $|C(S_i)| \geq \delta \times |C(L_i)|$ then we call $S$ has *acceptable coverage*, AC, property. Under this condition, we call $a$ the $AC$ parameter of $S$ and show it by $a_{AC}(S)$. In this definition $\delta$ is the degradation coefficient.

**4- Minimal Set** Another important property of the selected sensors is that in each time step, the selected sensor-set should be minimal. In other word, no subset of the selected sensor-set should have coverage equal to the whole selected sensors.

Since there might be more than one sequence of selected sensors set with mentioned properties our goal is to select the one which has the maximum of $a_{IC}$ as well as the maximum of $a_{FC}$.

Figure 2 shows an example for the behavior of the mentioned parameters.

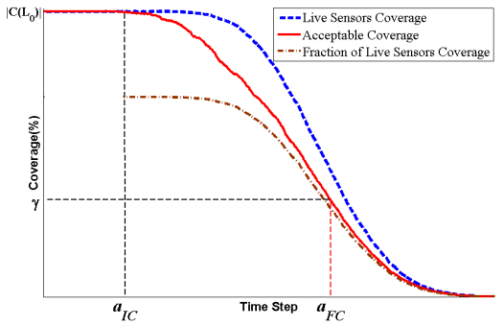

Fig. 2. Typical behavior of sensor network.

In Fig. 2, the upper curve belongs to the coverage of all live sensors. The bottom curve is plotted with multiplying the upper curve by the degradation coefficient, $\delta$. The goal is not to get coverage below that of the bottom curve. The middle curve is an example of an attempt to stay in between the two boundaries and hold the AC property. It can be seen in this figure that after the time step $a_{IC}$ it is not possible to achieve the initial available coverage of $C(L_0)$. Hence, the available coverage is $C(L_i)$ for $i \leq a_{IC}$. If we were to achieve this $C(L_i)$ then we have to turn on all of the sensors which results in a very rapid diminishing of the resources. Hence, after $a_{IC}$ the role of $\delta$ is to allow graceful degradation of the coverage area by controlling the behavior of the algorithm. If a graceful degradation of the resources were to be achieved then after $a_{IC}$ the coverage set of less than $|C(L_0)|$ will be reached. The best achievable coverage will be a value greater than or equal to $\delta \times |C(L_i)|$. Selecting sensors to give this much coverage will spare some of the sensors until the coverage of $\gamma \times |C(L_0)|$ is encountered at time $a_{AC}$ when the network is consider as dead. In Fig. 1 the upper graph shows $|C(S_i)|$ at each time step but if we were to achieve this coverage instead of $|C(L_0)|$ the network would have collapsed very quickly after $a_{IC}$.

Achieving the maximum coverage has higher priority to providing an acceptable coverage set. However, when the maximum possible coverage cannot be achieved, one may prefer to have an acceptable level of coverage, keeping the network in operation with the longest possible duration. In the coming sections, a new approach is proposed to achieve coverage, based on mentioned properties of the network.

## III. PROPOSED APPROACH

In this section, the proposed solution of the sensor coverage and lifetime problem is described in details. This solution is constructed by different stages which are shown in Fig. 3.

The first step of the proposed method is to randomly distribute a certain number of camera sensor nodes over the sensing area. This step is followed by camera calibration stage. Camera calibration is required in order to inform the base station of the geographical position and the angular orientation of each sensor.

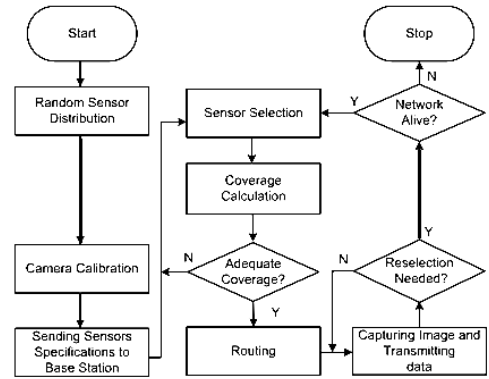

Fig. 3. Block diagram of proposed method.

When a sensor transmits a captured image, the base station knows exactly which part of the airspace is photographed by that specific sensor. Therefore, it is possible for the base station to select a number of sensor nodes based on some criteria. Sensor selection can be performed based on different algorithms with the aim of coverage optimization. In the following section, novel algorithms are proposed for this purpose.

When the adequate coverage is reached, the next step is to find appropriate routes between selected sensors and the base station. This step is performed during the routing stage. In this paper, for routing, we employ a modified version of Multicast Ad-hoc On-demand Distance Vector (MAODV) [15]. Data transmission starts after the routing stage is finished. During the data transmission, selected sensor nodes, transmit their data to the base station through the pre-constructed routes. However, due to limited source of energy, after a while, some of the nodes fail which result in the failure of some of the routes. Under these conditions, the base station has to re-select sensor nodes and hence, sensor selection and routing stages must be repeated while the network is alive.

## IV. PROPOSED SELECTION ALGORITHMS

In this section, a prioritized function is proposed with which several methods with different properties could be developed. Based on the proposed function and among different probable methods we develop two of them. The performance of these strategies is evaluated in Section V through simulations.

Algorithm 1 illustrates details of our approach. It is assumed that this algorithm is run at $j^{th}$ time step of network operation. In the first line, to see if the network is alive or not, validity of FC property is checked. It means that if the coverage of live sensors is less than a predefined threshold then network is assumed to be useless. After that, the goal is to select sensors with adequate coverage. If $C(L_0)$ can be achieved, then adequate coverage is $C(L_0)$, otherwise, when the current maximum coverage is less than $C(L_0)$ then AC should be achieved and adequate coverage is at least equal to $\delta \times |C(L_i)|$. Therefore, the adequate level of coverage, adaptively changes with the network conditions. The set $S_j$ contains sensors which provide adequate coverage. On the other hand $S_j$ is the set of selected sensors at $j^{th}$ time step. At every iteration of this algorithm, a sensor is selected based on its priority and added to $S_j$. Parameter $k$ is the index of a sensor based on the employed priority function, $Priority_i$. In line 5 it is checked to see if the maximum possible coverage can be achieved or not. If the current coverage is less than the whole coverage of live sensors right at the beginning of network's operation, $C(L_0)$, then AC must be considered ($\delta = \delta_1 < 1$). Otherwise, the IC is desirable ($\delta = 1$). Eventually, the algorithm forms the set $S_j$ of the selected sensors.

In the 10$^{th}$ line of the algorithm it is assumed that to prioritize $i^{th}$ sensor, a function $f$ can be utilized which depends on the remaining energy $E_i$, amount of overlap OV, and the coverage of the live sensors at $j^{th}$ time step, $C(L_j)$ In this function, OV is a set of points in the target region composed of the intersection of at least two sensors coverage sets. On the other hand if $X$ is an arbitrary set of sensors, then we define OV(X) to be $OV(X) = \{y | y \in C(\{m\}) \cap C(\{n\}), m, n \in X\}$.

After assigning a priority to each live sensor, the sensor with the highest priority among the unselected ones is selected at line 12 and its index is stored in k. Then, the sensor is added to the list, $S_j$, containing the selected sensors as mentioned in line 13.

While we experimented with many different combinations of $E_i$, OV and $C(L_j)$, in this paper two of them are presented. These functions and also their corresponding algorithms are named maximum coverage area (MA) and minimum lifetime and minimum overlap (MLMO). In the MA selection method, sensors are selected based on their coverage area. In other words the function $f$, in the 10$^{th}$ line of the algorithm, is defined by $f = |C(\{i\})|$. In this method, a list of sensors is sorted based on their coverage area. Live sensors with the greatest coverage areas are selected from top of the list.

```
BEGIN
01    IF |C(L_j)| < γ|C(L_0)|
02         EXIT
03    S_j = ∅, k = 1, δ = 1
04    IF |C(L_j)| < |C(L_0)|
05         δ = δ_1
06    WHILE (|C(S_j)| < δ × |C(L_j)|)
07    BEGIN
08         FOR i ∈ L_j
09         BEGIN
10            Priority_i = f(E_i, OV({i} ∪ S_j), C(L_j))
11         END
12         I_k = argmax{Priority_i}
                  i∈L_j−S_j
13         S_j = S_j ∪ {I_k}
14         k = k + 1
15    END
16    FOR i ∈ S_j
17         IF |C(S_j)| = |C(S_j − {i})|
18            S_j = S_j − {i}
19    END
END
```

Algorithm 1: Sensor selection algorithm.

In the MA selection method, the coverage area of each sensor is considered separately. However, an important phenomenon that should be considered in the sensor selection is the overlap between the coverage areas of two sensors which are visually neighbors. Selection of two sensors with large visual overlaps should be avoided. This phenomenon is considered in the MLMO method. A method, which can consider the amount of overlap, is the formation of the OV set for each sensor and assigning higher priority to a sensor with smaller overlap, |OV|. Since many sensors might have similar overlap values, considering this parameter alone is not appropriate especially at the first step of the sensor selection process. It means that it is more appropriate to combine overlap with other criteria such as the remaining energy and/or the coverage of a sensor. Therefore in the MLMO method, in addition to overlap, we also consider the remaining battery power of each sensor. The function that we apply to prioritize the sensors based on MLMO is defined by by $f = 1/(|E_i| \times (|OV(\{i\} \cup S_j)| + 1))$. Since overlap of a sensor might be zero, to avoid division by zero, in the denominator, 1 is added to $|OV(\{i\} \cup S_j)|$.

The proposed methods are evaluated by simulation and their results are presented in the following section.

## V. SIMULATION RESULTS

A large variety of simulations has been performed using a broad range of parameter values. In this section, some of the results are presented to show the relative performance of the proposed algorithms.

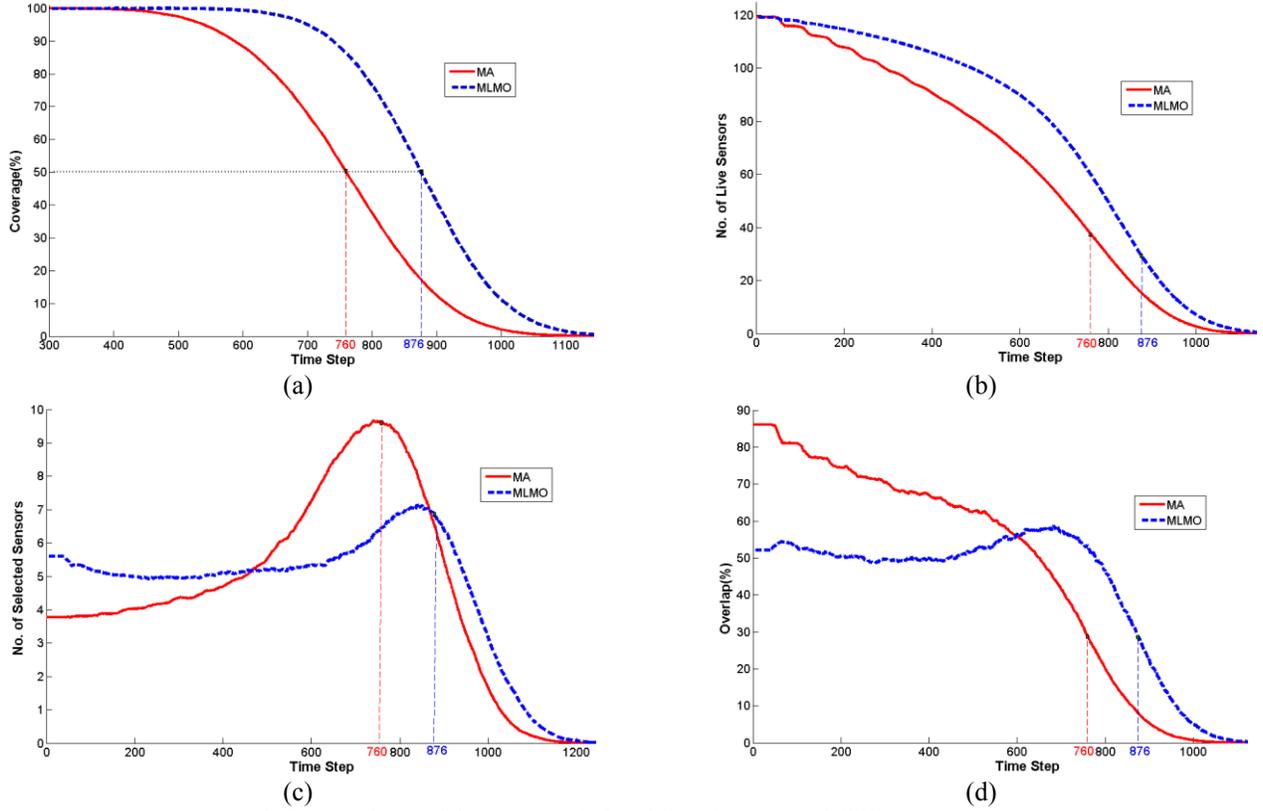

Fig 4. Comparison of the proposed algorithms in terms of different parameters

In order to do the simulation a number of camera sensors are distributed over the simulation area. The initial number of sensor nodes is $|L_0|$ that are distributed over an $L_x \times L_y$ rectangular area in a uniform random manner. The initial energy of each camera is randomly selected between $E_{min}$ and $E_{max}$. Initial values of the simulation parameters are listed in Table I

Table I Simulation Parameters

| Parameter | Value | Parameter | Value |
|---|---|---|---|
| $E_{min}$ | 1000 | $\gamma$ | 50% |
| $E_{max}$ | 1300 | $L_x$ | 100 |
| $\delta$ | 0.95 | $L_y$ | 100 |
| $|L_0|$ | 200 | $r$ | 25 |

Comparison of the results of the proposed algorithms is presented in Fig. 4(a) to Fig. 4(d).

In order to evaluate the proposed algorithms five criteria are used. These criteria are the number of live sensors, $L_i$, number of the selected sensors, $S_i$, and coverage percentage, defined by $|C(S_i)|/|C(L_0)| \times 100$. Also we use percentage of overlap, defined by $\frac{|OV(S_i)|}{|C(S_i)|} \times 100$ in our comparisons. In addition to the above parameters, we use time-coverage defined by the integral of sensors' coverage sets over network lifetime to compare different algorithms. Since our goal is to maximize the network lifetime, the algorithm that maximizes the time-coverage is preferable.

Percentage of the coverage during network lifetime is shown in Fig. 4(a). Since both algorithms have produced complete coverage before 300[th] time step, for better illustration, these curves are shown only for time steps greater than 300. As it is shown in the Fig. 4(a), MLMO prolongs network lifetime while it has the higher coverage too. It is assumed that once the coverage of an algorithm falls below $\gamma = 50\%$ of the total coverage, the network has failed. The point of such failure is specifically illustrated in Fig. 4(a) for each of the algorithms. Even though the results are shown until the entire sensors die but the comparisons are performed based on the 50% points. For example, for the MA algorithm, after 1250 time steps none of the sensors is alive but we are only interested in the network as long as enough sensors are alive to give at least half of the complete coverage. The half-coverage lifetime for this algorithm is 760 time steps which is marked on the corresponding graph.

Fig. 4(b), shows the number of live sensors for both algorithms. Although each algorithm chooses subsets of sensors that are different from the others, these algorithms operate with the same number of live sensors during a few initial steps of their lifetimes. It is because that the initial energies of all of the sensors are higher than a certain threshold that is mentioned in Table I. Since it takes some time for a number of sensors to die, the number of live sensors differs for each algorithm.

Number of the selected sensors at each time step of the network's operation is shown in Fig. 4(c). In the MA algorithm and at the first step of the network lifetime, since there is overlap among sensors coverage sets, lower number of sensors need to be selected. The initial selected sensors have large coverage areas which result in larger overlaps. Later on, the selected sensors with smaller coverage areas produce smaller overlap regions. The selection process is repeated until the remaining live sensors fail to realize the initial coverage (IC). At that point, the sensor selection is performed based on the proposed algorithms to cover an acceptable part of the targeted region. Also MLMO behaves similarly, except that its selected sensors at the initial steps of the network operation are more than that of MA's. It's because that the sensors with higher coverage areas are selected sequentially in the MA while in MLMO their overlaps are also considered in the selection process. Therefore a sensor with lower coverage than that of MA's can be selected in MLMO which results in the increase of the number of the selected sensors.

Figure 4(d) illustrates percentage of overlap of each algorithm. Since in the selection phase of MLMO the amount of overlap is considered, it is expected that these algorithms produce less overlap in comparison with MA.

In Table II, the effects of the proposed algorithms are compared in terms different metrics. Based on the simulation parameters mentioned in Table I, MLMO performs better than MA for all of the metrics.

Table II: Comparison of proposed algorithms in terms of lifetime, coverage, and number of involved sensors.

| Algorithm | Time-coverage product | Lifetime | Overlap | No. of selected sensors | No. of live sensors |
|---|---|---|---|---|---|
| **MA** | 58.9 | 760 | 40.7 | 4.48 | 55.7 |
| **MLMO** | 61.4 | 876 | 32.8 | 3.98 | 59.9 |

## VI. CONCLUSION

In this paper the use of visual sensor networks for observation of a designated aerial territory was studied. Due to the fact that the sensor nodes used in these networks rely on non-rechargeable energy sources, the goal of the sensor selection algorithms is to maintain complete visual coverage of the area for the longest period of time. To form a complete coverage of the target plane, a group of sensors must be selected from an initially large number of randomly distributed sensors. In order to select a subset of sensors that can cover our desired area we assigned each sensor a priority based on its remaining energy, its coverage, and the overlap area. Among different probable combinations of these parameters that can produce different priority functions, we used two of them and through simulations we showed that these parameters have important roles in network lifetime. These priorities were coverage; overlap and remaining energy; and symmetric difference between each sensor coverage and previously covered area. Also, during the operations of the algorithms we considered two stages. In the first stage of the operation, sensors were able to maintain complete coverage of the designated target plane. Hence, minimum sensors, with highest priorities, were selected to produce the desired complete visual coverage. In the second stage, due to the termination of a number of sensors, complete coverage could not be maintained and a percentage of that coverage was achieved. This helped the elongation of the lifetime of the network.

REFERENCES

[1] Lobaton, E. J., Ahammad, P., & Sastry, S. (2009, April). Algebraic approach to recovering topological information in distributed camera networks. In Proceedings of the 2009 International Conference on Information Processing in Sensor Networks (pp. 193-204). IEEE Computer Society.

[2] Charfi, Y., Wakamiya, N., & Murata, M. (2009). Challenging issues in visual sensor networks. IEEE Wireless Communications, 16(2).

[3] Rekleitis, I., New, A. P., & Choset, H. (2005). Distributed coverage of unknown/unstructured environments by mobile sensor networks. In Multi-Robot Systems. From Swarms to Intelligent Automata Volume III (pp. 145-155). Springer, Dordrecht.

[4] Akyildiz, I. F., Melodia, T., & Chowdhury, K. R. (2007). A survey on wireless multimedia sensor networks. Computer networks, 51(4), 921-960.

[5] Lobaton, E., Vasudevan, R., Bajcsy, R., & Sastry, S. (2010). A distributed topological camera network representation for tracking applications. IEEE Transactions on Image Processing, 19(10), 2516-2529.

[6] X. Liu, P. Kulkarni, P. Shenoy, and D. Ganesan, "Snapshot: a self-calibration protocol for camera sensor networks," 3 rd Int. Conf. on Broadband Communications, Networks and Systems, pp. 1-10, 2006.

[7] C. Kandoth and S. Chellappan, "Angular mobility assisted coverage in directional sensor networks," in Int. Conf. on Network-Based Information Systems (NBIS), USA, 2009, pp. 376-379.

[8] Y. Wu, J. Yin, M. Li, Z. En, and Z. Xie, "Efficient algorithms for probabilistic k-coverage in directional sensor networks," in Int. Conf. on Intelligent Sensors, Sensor Networks and Information Processing(ISSNIP), Australia, 2008, pp. 587-592.

[9] H. Ma, X. Zhang, and A. Ming, "A coverageenhancing method for 3D directional sensor networks," in IEEE INFOCOM, Brazil, 2009, pp. 2791-2795.

[10] Hooshmand, M., Soroushmehr, S. M. R., Khadivi, P., Samavi, S., & Shirani, S. (2013). Visual sensor network lifetime maximization by prioritized scheduling of nodes. Journal of Network and Computer Applications, 36(1), 409-419.

[11] Khadivi, P., Todd, T.D., Samavi, S., Saidi, H., Zhao, D., "Mobile Ad Hoc Relaying for Upward Vertical Handoff in Hybrid WLAN/Cellular Systems", Elsevier Ad Hoc Journal, Vol. 6, No. 2, pp. 307-324, April 2008.

[12] Khadivi, P., Samavi, S., Saidi, H., & Todd, T. D. (2006, June). Handoff in hybrid wireless networks based on self organization. In Communications, 2006. ICC'06. IEEE International Conference on (Vol. 5, pp. 1996-2001). IEEE.

[13] S. Soro and W. Heinzelman, "Camera selection in visual sensor networks," in IEEE Conf. on Advanced Video and Signal Based Surveillance(AVSS), U.K, 2007, pp. 81-86.

[14] Zamora, N. H., & Marculescu, R. (2007, September). Coordinated distributed power management with video sensor networks: Analysis, simulation, and prototyping. In Distributed Smart Cameras, 2007. ICDSC'07. First ACM/IEEE International Conference on (pp. 4-11). IEEE.

[15] Royer, E. M., & Toh, C. K. (1999). A review of current routing protocols for ad hoc mobile wireless networks. IEEE personal communications, 6(2), 46-55.